\documentclass{article}

\usepackage{natbib}
\usepackage{amsmath,amssymb}

\usepackage{nameref}
\usepackage[colorlinks = true,
            linkcolor = blue,
            urlcolor  = blue,
            citecolor = blue,
            anchorcolor = blue]{hyperref}

\usepackage{bbm}
\usepackage{graphicx}

\title{Bayesian Genome- and Epigenome-wide Association Studies with Gene Level Dependence}

\author{Eric F. Lock  \\
	   Division of Biostatistics, University of Minnesota\\  Minneapolis, MN 55455, U.S.A
	   \and 
	   David B. Dunson\\
	   Department of Statistical Science, Duke University, \\  Durham, NC 27708, U.S.A.
	   }

\date{}
\begin{document}

\maketitle

\begin{abstract}
High-throughput genetic and epigenetic data are often screened for associations with an observed phenotype.  For example, one may wish to test hundreds of thousands of genetic variants, or DNA methylation sites, for an association with disease status.  These genomic variables can naturally be grouped by the gene they encode, among other criteria.  However, standard practice in such applications is independent screening with a universal correction for multiplicity.  We propose a Bayesian approach in which the prior probability of an association for a given genomic variable depends on its gene, and the gene-specific probabilities are modeled nonparametrically.  This hierarchical model allows for appropriate gene and genome-wide multiplicity adjustments, and can be incorporated into a variety of Bayesian association screening methodologies with negligible increase in computational complexity.   We describe an application to screening for differences in DNA methylation between lower grade glioma and glioblastoma multiforme tumor samples from The Cancer Genome Atlas.  Software is available via the package {\tt BayesianScreening} for R: \href{https://github.com/lockEF/BayesianScreening}{\tt github.com/lockEF/BayesianScreening}.\\
\end{abstract}

\section{Introduction}

	Several technologies that are used for genomic research measure data that are high-throughput and genome-wide.  These data may be genetic or epigenetic.  Technologies that measure genetic data include single nucleotide polymorphism (SNP) arrays, whole-exome sequencing, and whole-genome sequencing;  technologies that measure epigenetic data include DNA methylation bisulphite arrays or bisulphite sequencing,  and chromatin immunoprecipitation sequencing (ChIP-seq).  These technologies all measure hundreds of thousands of variables, each of which can be mapped to a location on the genome.  In this article we use the general term ``marker" to refer to any such variable.    
	
	A recurring objective in genomic research is to test each marker for an association with a given phenotypic trait, such as disease status.  These are commonly conducted in a frequentist framework, where a p-value for the null hypothesis of no association is calculated independently for each marker.  Several thousand such studies have been conducted for genetic associations alone \citep{welter2014}.  While these studies have revealed several important biomarkers, they have also been criticized for lack of power and lack of reproducibility  \citep{visscher2012}.  The reliance on p-values and binary conclusions may be partly responsible for these criticisms.  P-values are a poor proxy for our degree of confidence that a true association exists, because they depend on the power of the test \citep{stephens2009}.  Furthermore, standard corrections for multiple comparisons that control the family-wise error rate or false discovery rate for a single study typically require exorbitant effect sizes, leaving most associated markers undetected \citep{park2010}.
	
	As an alternative to frequentist-based approaches, several methodologies have been developed to screen for genome-wide associations in a fully Bayesian framework (for a review see \citet{stephens2009}), and these are increasingly used in practice.  Bayesian approaches directly compute the posterior probability that a marker is associated with a given trait, under a full probabilistic model for both the null and alternative hypotheses.  This provides a straightforward and intuitive framework for meta-analyses that combine results from multiple studies \citep{verzilli2008, wen2014}.  Moreover, Bayesian approaches provide a natural framework for borrowing information across multiple related markers within a single study to compute more well-informed and accurate weights of evidence in the form of posterior probabilities.  Importantly, Bayesian techniques that combine multiple related tests need not treat the null and alternative hypotheses asymmetrically; this is in contrast to frequentist approaches to the multiple comparisons problem that typically require larger effect sizes for the alternative as the number of tests grows.  
 
Despite their potential flexibility for borrowing information, standard practice for Bayesian genome-wide testing is to screen each marker independently.  This involves specifying a prior probability for association at each marker \citep{stephens2009}, or effectively fixing the prior probability at $0.5$ and considering the Bayes factor for each marker \citep{wakefield2009,xu2012}.  Alternatively, the prior probability of association at each marker can be treated as unknown (with, for example, a $Beta(a,b)$ prior distribution) and inferred during posterior computation \citep{scott2010, lock2015}.  However, this approach still relies on the over-simplified premise that the probability of association is the same for all markers.  
 
In this article, we describe a computationally scalable and widely applicable approach to inferring null probabilities that depend on the genomic location of each marker.  Specifically, the prior probability of association for a given marker depends on the gene it encodes.  The gene-specific probabilities are modeled with a nonparametric distribution that allows for appropriate genome-wide adjustments for multiplicity.  We demonstrate how this approach can dramatically improve posterior accuracy and interpretation when there is gene-level dependence among tests.   

We apply our approach to an epigenome-wide association study of cancerous brain tumors that develop from astrocyte cells.  Specifically, we use DNA methylation data from the Illumina HumanMethylation450 array to compare methylation profiles between lower-grade astrocytoma and glioblastoma multiforme samples from The Cancer Genome Atlas.   These data include methylation measurements at $294,093$ genomic sites that map to $24,358$ different genes.  We apply our gene-level dependence model in conjunction with a previously described method for screening for differential distribution between groups in methylation array data based on shared kernels \citep{lock2015}.  Our analysis reveals systematic differences in methylation distribution at a large number of genomic sites, and the proportion of sites with differential methylation varies substantially between genes. 

\subsection{Gene-wise Association Tests}
	
Many methods have been developed that combine multiple markers within a single gene to test for an association at the gene level.  For example, there is a wide body of literature on methods that aggregate genetic variants within a gene, via a direct sum or a regression model, to obtain a p-value for the null hypothesis that the gene has no association with the given phenotype \citep{pan2014,wu2011,liu2010}.  Similarly, there are methods that combine methylation markers within a given gene (or region) to obtain a composite p-value \citep{wang2012}.  These methods can substantially increase power if many markers within a gene have a weak association that cannot be detected independently \citep{wojcik2015}, and also reduce the number of overall tests for multiplicity correction.  However, aggregating at the gene level may miss important marker-specific effects;  for example, different mutations within the same gene can have very different phenotypic consequences \citep{rowntree2003}.

In a Bayesian framework, \citet{wilson2010} describe a genome-wide model for the association of genetic markers with an observed phenotype, in which the Bayes factor for model inclusion can be computed at the marker or gene level.  In their implementation each marker has the same prior probability of association $p$ genome-wide, with hyperprior $p \sim Beta(a,b)$;  a gene is considered associated with the observed phenotype if any marker within the gene is associated.  Alternatively, \citet{ruklisa2015} describe a class of Bayesian approaches to rare variant association testing in which the prior probability that a given marker is associated depends on the gene it encodes.  For their approach the gene-specific probabilities are estimated independently based on training data, with no borrowing of information across the genes.  Nevertheless, they illustrate that gene-specific probabilities can outperform genome-wide approaches.   
 
In this article we describe a flexible compromise between genome-wide and gene-specific priors for marker associations. 

\section{Model}
\label{model}

Here we describe our hierarchical model for gene-specific probabilities in general, to convey its applicability to a wide variety of data types and Bayesian models for association.  Details specific to the methylation screening example in Section~\ref{app} are given in Appendix~\ref{appendix}.

Suppose data are collected for $M$ genetic or epigenetic markers from $N$ individuals, where each marker maps to one of $G$ genes.  Let $M_g$ be the number of markers that map to gene $g$, so that $M=\sum_{g=1}^G M_g$.  Let $X_{gmn}$ denote data for marker $m$ in gene $g$ ($g \in 1,\hdots,M_g$) for individual $n$, and let $Y_n$ define a phenotypic response for individual $n$.  

Let $H_{0, gm}$ define a probabilistic model of no association with $Y$ for marker $m$ in gene $g$, and $H_{a, gm}$ define the alternative model of association.  The posterior probability of the null for the given marker, $P(H_{0, gm} \mid X,Y)$, is 
\[\frac{P(H_{0, gm}) P(X,Y \mid H_{0, gm})}{P(H_{0, gm}) P(X,Y \mid H_{0, gm})+P(H_{a, gm}) P(X,Y \mid H_{a, gm})}.\] 
Under our proposed model, prior probabilities are equal within a gene:
\[p_g = P(H_{0,gm})  \, \, \, \mbox{for} \, \, \, m=1,\hdots, M_g. \]
We use a nonparametric hyperprior to infer the gene-level prior probabilities $\{p_g\}_{g=1}^G$ and borrow information across the genes.  Specifically, the distribution of the $p_g's$ is a Dirichlet process \citep{ferguson1973} with a $\mbox{Beta}(a,b)$ base distribution and concentration parameter $\alpha$: $p_g \sim \mbox{DP}(\mbox{Beta}(a,b), \alpha).$ 
Under this framework, each $p_g$ is drawn from a theoretically infinite number of realizations $\theta_h$ from Beta$(a,b)$, with corresponding probability weights $\pi_h$:
\[p_g \sim \sum_{h=1}^\infty \pi_h \delta_{\theta_h},\] 
where $\delta_{\theta_h}$ is a point mass at $\theta_h$.  A consequence of this model is clustering of the genes, as values $\theta_h$ with larger weights $\pi_h$ will correspond to the probability for several genes.  This clustering property is useful for interpretation (e.g., to identify gene sets) but our primary motivation for using the Dirichlet process is to provide a sufficiently robust and flexible hierarchical distribution for the $p_g's$.

The concentration parameter $\alpha$ controls the dispersion of the weights $\pi_h$ and, hence, influences the sizes of the gene clusters.  As $\alpha \rightarrow 0$ a single realization will correspond to all genes (e.g., $p_1=\hdots=p_G=\theta_1$); hence, the limit is equivalent to a genome-wide correction for multiplicity.  As $\alpha \rightarrow \infty$ each gene will have its own realization (e.g., $p_1=\theta_1, \, \, p_2=\theta_2, \,\, \hdots)$;  hence, the limit is equivalent to a separate, independently estimated probability for each gene.  In practice we find that fixing $\alpha$ as a small positive value, such as $\alpha=1$, allows for sufficient posterior flexibility between these two extremes.   

It is also informative to consider the choice of $a,b$ in the Beta base distribution.  In applications where a Beta$(a,b)$ distribution is used for a shared prior probability, fixing $a=1$ is common \citep{scott2010}.  Choosing $b=\lambda M$, provides a natural multiplicity adjustment, as the expected number of associated markers under the prior model is then $1/\lambda$ regardless of the number of markers $M$ \citep{wilson2010}.  This result extends to our context, as $E(p_g) = a/b$ and therefore the expected number of associated markers under the prior is 
\[\sum_{g=1}^G M_g E(p_g) = M \cdot \frac{a}{b}.\] 
However, philosophically there may be little reason for the probability of association at each marker to be negatively effected by the number of markers measured.  In practice we find that a simple uniform base distribution Beta$(1,1)$ allows for substantial flexibility and still performs well as a multiplicity correction under a global null.  

The parameter $p_g$ should not be interpreted as the overall probability of association for gene $g$.  Rather, it can be viewed as the inferred proportion of locations within the gene that are associated.  This is one approach to prioritize genes, but more importantly the $p_g's$ can improve the accuracy of posterior inference at the marker level.  

\section{Inference}
\label{comp}

Here we describe a general Gibbs sampling scheme to compute the full posterior under the gene-level prior model specified above.  This estimation approach is informative, illustrating how the marker parameters, gene parameters, and global parameters relate to each other.  Fundamentally, the algorithm proceeds by  sampling from the posterior of each marker, then updating the gene-specific probabilities and their corresponding Dirichlet process parameters. 

 We use the constructive stick-breaking representation of the Dirichlet process \citep{sethuraman1994} to sample from its full conditional distribution.  That is, the probability weights $\pi_h$ are generated by $\pi_h = V_h \prod_{l<h} (1-V_l)$, where $V_h \overset{iid}{\sim} \mbox{Beta}(1, \alpha)$.  In practice we truncate the infinite mixture by a large integer $H$, and perform blocked Gibbs sampling \citep{ishwaran2001}.  Thus, letting $C_g$  define the cluster index for gene $g$ ($p_g = \theta_{C_g}$), $C_g \in \{1,\hdots,H\}$.  The weights $\pi_h$ usually decrease quickly to very small values, and thus the effect of truncation is negligible.         

Assuming the marginal likelihoods under the null and alternative models can be computed for each marker, sampling from the full conditionals proceeds as follows:  
\begin{enumerate}
\item Designate null markers $H_{0,gm}$ for $g=1,\hdots,G$, $m=1,\hdots,M_g$. The conditional probability of the null, $P(H_{0,gm} \mid X,Y,p_g)$, is
\[\frac{p_g P(X,Y \mid H_{0, gm})}{p_g P(X,Y \mid H_{0, gm})+(1-p_g) P(X,Y \mid H_{a, gm})}.\]
\item Allocate indices $C_g$ for $g = 1,\hdots, G$:
\[P \left(C_g=h \mid \theta_h, \{H_{0,gm}\}_{m=1}^{M_g}\right) \propto \pi_h  \theta_h^{S_g} (1-\theta_h)^{M_g-S_g}\] 
 for $h=1,\hdots,H$, where $S_g$ is the number of null markers in gene $g$, $S_g=\sum_{m=1}^{M_g} \mathbbm{1}(H_{0,gm})$.
 \item Update the weights $\pi_h$ for $h=1,\hdots,H$.  First, draw the stick-breaking weights $V_1,\hdots,V_{H-1}$.  The full conditional distribution of $V_h$ is  
\[\mbox{Beta} \left(1+\sum_{g=1}^G \mathbbm{1}(C_g=h), \alpha+\sum_{g=1}^G \mathbbm{1}(C_g>h)\right), \]
 with $V_H=1$.  Then set  $\pi_h = V_h \prod_{l<h} (1-V_l)$ for $h=1,\hdots,H$.
 \item Update the atoms $\theta_h$ for $h=1,\hdots,H$.  The full conditional  distribution of $\theta_h$ is
 \[\mbox{Beta} \left(a+\tilde{S}_h,b+\tilde{M}_h-\tilde{S}_h\right),\]
 where $\tilde{M}_h$ is the total number of markers in genes allocated to cluster $h$, and $\tilde{S}_h$ is the number of null markers:
\[\tilde{M}_h = \sum_{\{g: C_g=h\}} M_g \, \, \, , \, \, \, \tilde{S}_h = \sum_{\{g: C_g=h\}} S_g.\]
Set $p_g = \theta_{C_g}$ for $g=1,\hdots,G$.  
\end{enumerate}
 
Point estimates for the gene-level probabilities $p_g$ and marker posterior probabilities $P(H_{0,gm} \mid X,Y)$ can be obtained by averaging their draws over the sampling iterations.    
 
The marginal likelihoods under the null and alternative hypotheses in sampling step $1$ may not be feasible to compute directly.  If not, additional sampling steps can be incorporated to update model-specific parameters for each marker under $H_{0,gm}$ and $H_{a,gm}$.  Such an approach is used for the two-group methylation screening scenario described in Section~\ref{app}.      

%\subsection{Overview of method}
%We introduce a general model in which the prior probability of association for a given marker depends on its gene $g$.  The gene specific probabilities $p_g$ are modeled nonparametrically, using a Dirichlet process with a $\text{Beta}$ base distribution.  In contrast to independent estimation of the $p_g$'s, this hierarchical model allows for appropriate shrinkage of the inferred $p_g$'s toward genome-wide patterns, as illustrated in the simulations below.  Under this framework, and a given Bayesian model for association at each marker, we estimate the full posterior distribution by iteratively sampling from the posterior probability of association for each marker, then updating the gene-specific probabilities and their corresponding Dirichlet process parameters.  For methodological details see the \nameref{methods} section. 

\section{Simulation Study}
\label{sims}

Here we present a simulation study to illustrate the advantages of our hierarchical model for gene-level probabilities.  We compare our hierarchical approach detailed in Sections~\ref{model} and ~\ref{comp} with three other approaches for inferring marginal probabilities of the null at each marker:
\begin{itemize}
\item Separate estimation, in which a probability is inferred independently for each gene, and shared by all markers for that gene.  
\item Joint estimation, 	in which a probability is inferred globally and shared by all markers.
\item Simple estimation, in which the prior is fixed at $0.5$ for all markers.  This is equivalent to independently considering the Bayes factor for each marker.
\end{itemize}

For simplicity, we assume marker values are binary (e.g., representing presence of the minor allele for genotype data).  We simulate data for two groups, each with $100$ individuals ($N=200$).  For null markers, binary values are simulated under a common probability for both groups, where this probability is drawn from a uniform distribution.  For alternative markers, binary values are simulated under a different probability for each group, where these probabilities are drawn independently from a uniform distribution.  The Bayes factor for the null over the alternative for a given marker is then
\[\frac{\beta(1+s_1+s_2, 1+200-s_1-s_2)}{\beta(1+s_1,1+100-s_1)\beta(1+s_2,1+100-s_2)},\]     
where $\beta$ defines the beta function, and $s_1$ and $s_2$ are the number of individuals for which the marker is present in groups $1$ and $2$, respectively.  This is analogous to the prospective Bayes factor for SNP association testing introduced in \citet{balding2006}.  Data are simulated for $G=1000$ genes, where the number of markers within a gene $M_g$ is drawn from $\{2,3,\hdots,20\}$ with equal probability.  

We consider three different scenarios with dramatically different assumption on the distribution of null and alternative markers across the genes.  For each scenario, we show the inferred distribution of the gene-specific probabilities $p_g$ under the four methods considered.  We also compute the expected overall error in classifying null and alternative markers, as the average misclassification probability over all markers.  
%\[\mathbbm{1}(H_{0,gm}) P(H_{a,gm} \mid X,Y) +\mathbbm{1}(H_{a,gm}) P(H_{0,gm} \mid X,Y)\]  

\begin{figure}
\begin{center}
\includegraphics[width=0.75\linewidth]{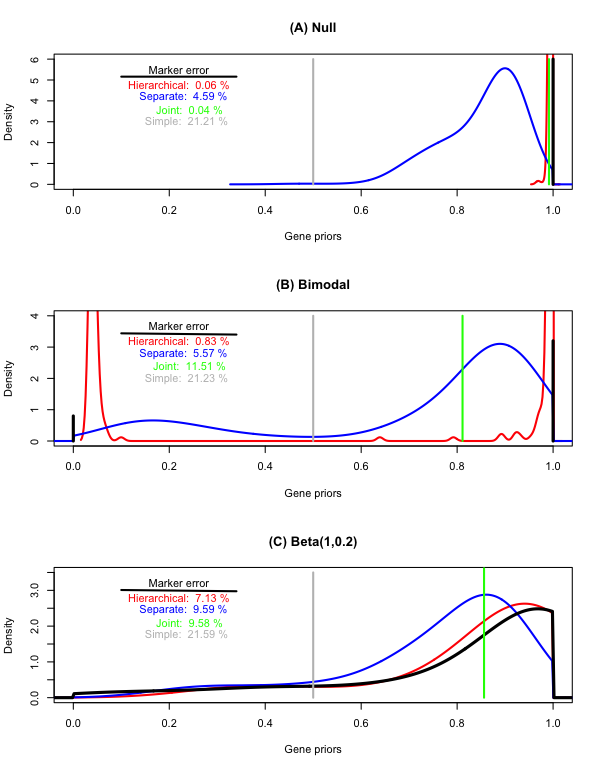}
\end{center}
\caption{Comparison of four approaches to inferring prior probabilities, under three simulation scenarios (A,B, and C).  Kernel density estimates of the resulting gene-specific probabilities are shown for continuous distributions;  discrete distributions are shown by vertical lines.  The distribution of the true gene-specific probabilities is colored black.  The expected overall error in classifying null and alternative markers is also shown for each method and each simulation scenario. }
\label{fig:sim}
\end{figure}

First, we simulate data where the null is true for all markers, to illustrate how the four methods perform as a multiplicity adjustment.  Results are shown in Figure~\ref{fig:sim}A.  The simple model with fixed prior probability of $0.5$ performs relatively poorly; in this and other simulations the average error in classifying markers independently is approximately $20\%$.  The joint and hierarchical models have negligible error, as they both borrow information globally to enforce appropriately high prior probabilities of the null.  The model with separately inferred priors for each gene does not perform as well, as its shift toward the null is relatively weak, especially for those genes with a small number of markers. 

Second, we simulate data from a bimodal distribution in which the majority of genes ($80\%$) are null for all markers, but for a subgroup of genes (20\%) the alternative is true for all markers.  Results are shown in Figure~\ref{fig:sim}B.  In this case the hierarchical model performs well, as it identifies both modes and allocates the appropriate genes to each mode.  The separate model performs better than the joint model, as the joint model does not account for the heterogeneity in the genes.  However, the separate model is not competitive with the hierarchical model, as again the gene-specific probabilities have substantial uncertainty and do not shrink toward the two modes if they are estimated independently.  

Third, we simulate the gene-specific probabilities from a Beta$(1,0.2)$ distribution, which has a majority of its mass near $1$ (corresponding to genes in which the vast majority of markers are null) but a long left tail.   Results are shown in Figure~\ref{fig:sim}C.    The joint and separate models perform similarly, as the joint model ignores the gene heterogeneity and the separate model exaggerates gene heterogeneity.  The hierarchical model serves as a flexible compromise between the two extremes, and closely approximates the true gene-specific probabilities.  

To compare the Bayesian methods above with frequentist methods for multiple hypothesis testing, we compute a p-value for the null using Fisher's exact test at each marker.  We consider the following multiplicity adjustments for these p-values:
\begin{itemize}
\item Separate false-discovery rate (FDR) corrections for each gene, using the Benjamini-Hochberg method \citep{Benjamini1995}.  
\item An overall FDR correction for each marker.
\item A two-step hierarchical hypothesis testing framework \citep{li2014} that uses the Hochberg \citep{hochberg1988} and Benjamini-Hochberg methods.  This method controls the overall FDR while allowing for dependence within sets of hypotheses.  In our context, a set corresponds to markers within a gene.    
\end{itemize}

 P-values and Bayesian posterior probabilities have fundamental differences in philosophy and interpretation, and are not directly comparable.  Nonetheless, for illustration we compare the various approaches above by considering standard thresholds on the posterior probability, p-value or FDR that are used to classify markers as null or alternative.  For Bayesian methods we use $0.5$ as a threshold on the posterior probability, and for the frequentist methods we use a significance threshold of $\alpha=0.05$.  The resulting misclassification rates under each simulation scenario are shown in Table~\ref{table:01}.  Under the null model the hierarchical Bayesian approach gives very low error, similar to overall FDR corrections. Under the two scenarios with alternative markers the Bayesian hierarchical model performs substantially better than frequentist multiplicity corrections, which are overly conservative.  

\begin{table}
\caption{Error in classifying null and alternative markers, using $0.5$ as a threshold for the posterior probability for Bayesian methods, and $0.05$ as an FDR or P-value threshold for frequentist methods using Fisher's exact test.  The average over $50$ replicate simulations is shown, and the resulting standard error is less than $0.05\%$ for all cells.}
\label{table:01}
\bigskip
\begin{center}
\begin{tabular}{l|c c c}
\hline
& Null & Bimodal & Beta $(1,0.2)$ \\
\hline 
\multicolumn{1}{c|}{\textbf{Bayesian}} & & &\\
Hierarchical &  $0.01\%$ & $0.07\%$ & $3.92\%$ \\
Separate & $0.50\%$ & $1.68\%$& $4.87\%$ \\
Joint & $0.01\%$ & $6.89\%$& $5.34\%$\\
Simple & $5.07\%$ & $8.72\%$ & $7.82\%$\\
\multicolumn{1}{c|}{\textbf{Frequentist}} & & &\\
Two-step FDR & $0.01\%$ &$7.73\%$ & $6.30\%$\\
Separate FDR & $0.23\%$ &$5.64\%$ & $5.41\%$\\
Overall FDR & $0.01\%$ &$7.15\%$ & $5.78\%$ \\
No correction &$2.91\%$ & $7.50\%$ & $6.49\%$\\
\hline
\end{tabular}%
\end{center}
\end{table}

\begin{figure}
\begin{center}
\includegraphics[width=0.90\linewidth]{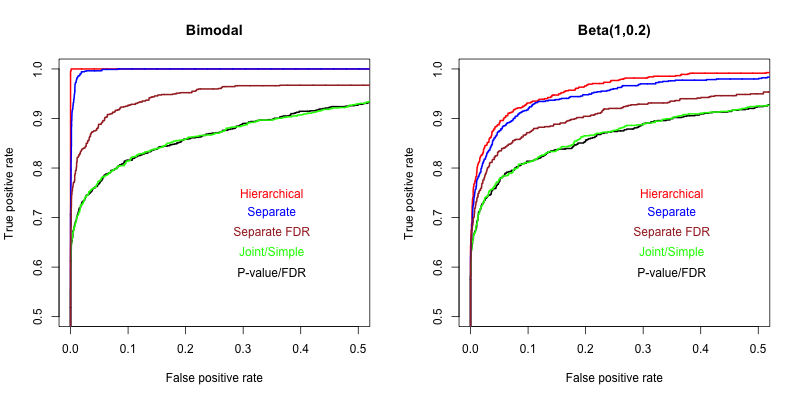}
\end{center}
\caption{ROC curves obtained by varying the threshold on the posterior probability, p-value, or FDR for various methods.  Curves are combined for the joint and simple Bayesian models, and for the uncorrected p-value and FDR, as the rank ordering of markers between these methods do not change.}
\label{fig:roc}
\end{figure}

Figure~\ref{fig:roc} gives receiver operating characteristic (ROC) curves showing the proportion of markers with false positive or true positive classification (where `positive' corresponds to the alternative) as the threshold on the posterior probability, p-value, or FDR is varied.  Results are shown for the bimodal and Beta simulations; the ROC curve for the null simulation is trivial, as there are no true positives.  In both cases the Bayesian hierarchical model has uniformly better classification performance than alternatives.  The ROC curves for separate estimation are close, indicating that the rank ordering of probabilities are similar despite the improved accuracy of the hierarchical model.    

A spreadsheet available online gives results for $99$ additional simulated datasets with varying sample size, number of genes, and distribution of gene-level probabilities.  These results demonstrate the robustness of the hierarchical model.   

\section{Application to LGG-GBM Methylation}
\label{app}
We implement our hierarchical gene-level prior model in a screen for differences in DNA methylation between lower grade gliomas (LGG) and glioblastoma multiforme (GBM) tumor samples that develop from astrocyte cells in the brain.  Methylation is an epigenetic phenomenon that occurs at cytosine-phosphate-guanine (CpG) dinucleotide sites in the genome.  Methylation is thought to play a significant role in LGG pathogenesis \citep{TCGA2015} and GBM pathogenesis \citep{brennan2013}, but the differences between the two tumor classes have not been well-characterized on a genome-wide scale.  Both tumors are heterogenous and typically fatal, but a more complete understanding of their molecular differences is important, as LGGs often progress to GBMs and GBM patients have a much shorter survival time.

We use data from the Illumina HumanMethylation450 array, for 128 astrocyte derived LGG samples and 130 GBM samples, from The Cancer Genome Atlas.  Measured CpG sites that have any missing data are removed, as are sites that map to intergenic regions.  After filtering, $294,093$ CpG sites remain, that map to $24,358$ distinct genes.  The number of sites in a gene ranges broadly from $1$ to $1017$.  Array values at each site are between $0$ (no methylation) and $1$ (fully methylated across all cells in the tumor sample).  

Several computational methods have been developed to screen for differential methylation levels between groups, based on array data \citep{jaffe2012,maksimovic2015} or sequencing data \citep{sun2014,feng2014,wu2015}.  However,  the focus on differential methylation levels between groups may miss other important differences between group distributions; for example, certain genomic regions have been shown to exhibit more variability in methylation, and hence greater epigenetic instability, among cancer cells than among normal cells \citep{hansen2011}.  

We test for differences in distribution between the LGG and GBM samples at each CpG site.  Our testing model is described in detail in \citet{lock2015}, where it is implemented on comparatively sparse methylation data ($\approx 20,000$ sites) with a global prior and shown to compare favorably to frequentist and Bayesian alternatives.  Briefly, the distribution at each site is modeled as a mixture of normal kernels, truncated between $0$ and $1$.  The kernels are shared across CpG sites, and thus capture shared patterns of multi-modality and skewness that are typical in methylation array data. Under the null the kernel mixing weights at each site are assumed to be the same between the two groups, and under the alternative they are different.  This provides a robust and consistent framework for testing differential distribution, and can identify important differences that are not captured by simply comparing mean methylation levels.  Furthermore, the method facilitates interpretation by modeling the full distribution, with uncertainty, for each class.  We incorporate hierarchical gene-level priors within the shared kernel testing model, and compute the full posterior via Gibbs sampling. 

\begin{figure}[h]
  \includegraphics[width=1\linewidth]{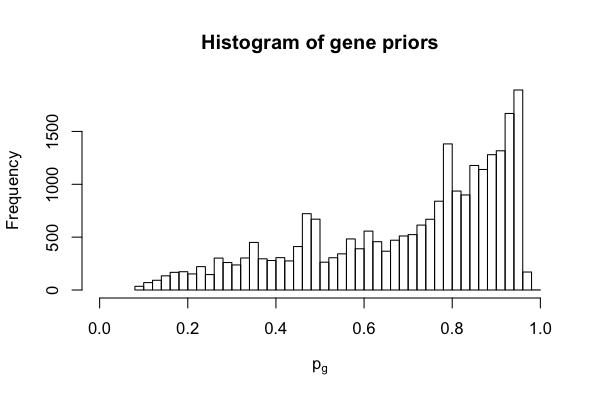}
\caption{Histogram of estimated gene-specific probabilities $p_g$.}
\label{fig:posthist}
\end{figure}

\begin{figure}[h]
\begin{center}
  \includegraphics[width=0.8\linewidth]{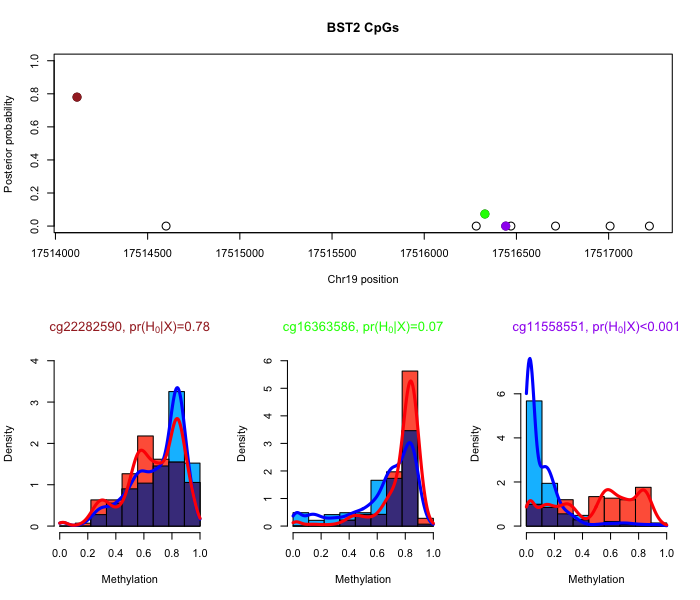}
 \end{center}
\caption{The top panel shows the estimated posterior probability of an association with GBM-LGG status for the 9 CpG sites measured in the gene BST2, with their corresponding genomic location.  The lower panel shows the estimated densities for the GBM (blue) and LGG (red) groups for three sites;  histograms of each group are shown, and their overlap is colored purple.}
\label{fig:BST2}
\end{figure}

The estimated gene-level probabilities are shown in Figure~\ref{fig:posthist}.  Their distribution resembles that in the simulation shown in Figure~\ref{fig:sim}C.  For the majority of genes the distribution between the two groups is inferred to be equal at most sites ($p_g \approx 1$).  However, there is a substantial left tail, corresponding to genes in which a large number of sites are inferred to differ between the two groups.  For illustration we focus on one such gene, BST2, which has $9$ measured CpG sites and an estimated gene-level probability of $p_g = 0.198$.  We select BST2 because it has been considered as a tangible target for immunotherapy in the treatment of GBM  \citep{etcheverry2010},  an independent comparison of GBM and normal samples found differences in BST2 methylation that correlate with gene expression \citep{wainwright2011}, and BST2 methylation may play a role in the pathonogenesis of other cancers \citep{mahauad2014}.  Figure~\ref{fig:BST2} shows the genomic location and posterior probability of group equality for the nine CpG sites in BST2, as well as group histograms and posterior densities for methylation at three sites.

\begin{figure}[h]
  \includegraphics[width=1\linewidth]{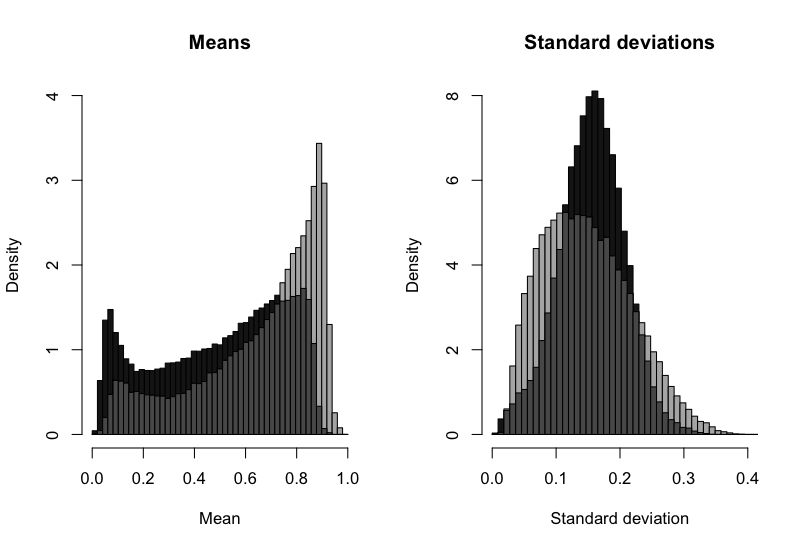}
\caption{Histograms of summary statistics for every CpG site with a posterior probability of the null less than $0.01$, computed separately for the GBM (dark gray) and LGG (light gray) groups.  Overlap between the two histograms is colored a neutral gray.  The left panel show the site means, the right shows the site standard deviations.  }
\label{fig:MeansSDs}
\end{figure}

Given the large number of CpGs and corresponding genes with differential methylation distribution, we also investigate differences at a macro level.  Figure~\ref{fig:MeansSDs} shows the site means and standard deviations within each group, for those sites with a posterior probability of equality less than $0.01$ (24.6\% of all CpGs).  Mean methylation levels at these sites are generally greater in the LGG samples than the GBM samples;  this is concordant with findings in a smaller comparison of 1536 CpG sites in 807 genes \citep{laffaire2011}.  The distribution of standard deviations is more curious, as LGG samples show a larger number of sites with either very high variability or very low variability in comparison to the distribution for GBM.

Details regarding the association model and Gibbs sampling algorithm for posterior computation are given in Appendix~\ref{appendix}.

\subsection{Validation} 

To asses the appropriateness of the hierarchical gene-level model, we consider the agreement of estimated gene-level probabilities and marker-level posteriors under cross validation.  Specifically, we randomly select $10,000$ CpGs to leave out, and compute gene-level probabilities using the remaining $284,093$ CpGs.  For each left out CpG, we measure the Kullback-Leibler divergence of its estimated gene-level probability under the reduced data from its CpG-level posterior probability under the full data.  This can be interpreted as the gain of information or degree of ``surprise" between a CpG's posterior probability and its gene-level prior \citep{lindley1956}.  We repeat this process using a separately estimated prior probability for each gene, a single inferred prior probability, and a prior probability of $0.5$ (corresponding to the separate, joint and simple models in the \nameref{sims}).  The hierarchical model yields the greatest agreement, with a mean Kullback-Leibler divergence of $0.451$; the separate model has a divergence of $0.482$, the joint model $0.560$, and the simple model $0.654$.       
 
\begin{figure}[h]
  \includegraphics[width=1\linewidth]{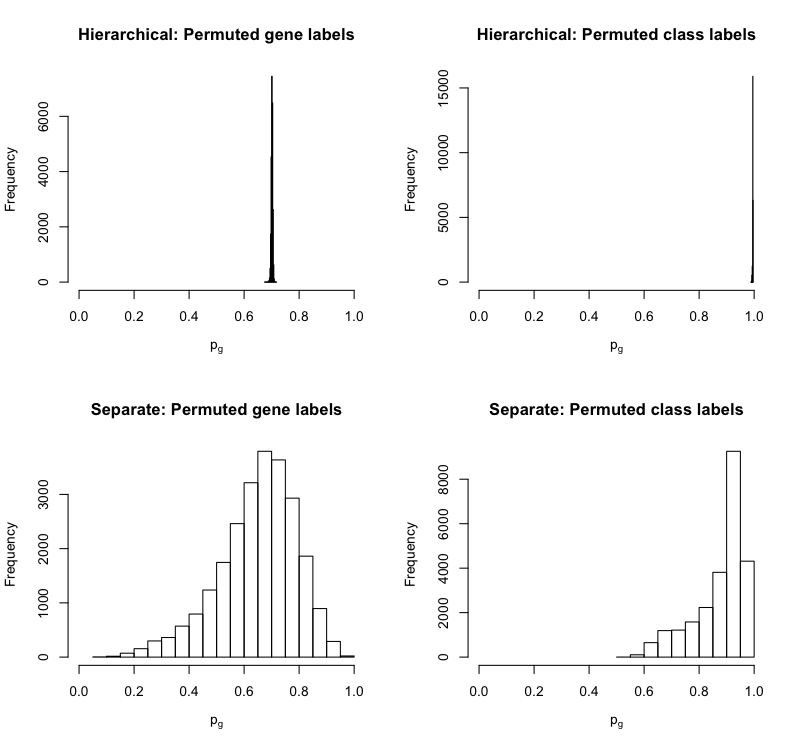}
\caption{Histogram of hierarchically estimated and separately estimated gene-level priors after randomly permuting gene labels (left column), and after randomly permuting class labels (right column).}
\label{fig:permhists}
\end{figure}

We also conduct two permutation studies, to further assess the appropriateness and flexibility of our gene-level model.  First, we randomly permute the gene labels for each marker, so that there is no true gene-level dependence.  The subsequent posterior means for the gene level prior probabilities $p_g$ are shown in the top-left panel of Figure~\ref{fig:permhists}.  The estimates converge appropriately to a single global probability near $0.72$, in sharp contrast to the relatively dispersed estimates using the true data in Figure 2 of the main article.  Second, we randomly permute the class labels but maintain the true gene labels, to generate a dataset with a global null but gene-level dependence.  The subsequent posterior estimates are shown in the top-right panel of Figure~\ref{fig:permhists}, and cluster very close to $1$.  In fact, all $294,093$ estimated site-specific posterior probabilities of the null are greater than $0.5$.  Together, these results demonstrate that the hierarchical gene-level model appropriately shrinks gene-level priors toward a global pattern. 

The estimated hierarchical gene-level probabilities closely approximate a single joint prior probability in Figure~\ref{fig:permhists}, where a joint prior is appropriate, but not for the true data.  Thus, we conclude that a single joint model is an over-simplification for these data.  We also compare the hierarchical gene-level prior probabilities under permutation with separate, independently estimated gene-level probabilities.  The separately estimated probabilities are shown in the bottom row of \ref{fig:permhists}.  These have a lot of variability under both permutation scenarios, illustrating how independent consideration of the genes sacrifices accuracy by exaggerating gene effects.     

\section{Discussion}
Borrowing information and incorporating prior knowledge in a principled and computationally feasible way is an important challenge for Bayesian genome- and epigenome-wide screening methods.  Here we present a flexible and generally applicable hierarchical model for inferring gene-specific probabilities, which may be extended in several ways.  Under our prior all markers within a gene have an equivalent probability of association.  The incorporation of other marker-level information, such as gene promoter status for DNA methylation \citep{weber2005} or functional annotation (e.g., synonymous vs. non-synonymous) for genotype markers \citep{kichaev2014,ruklisa2015}, may improve posterior precision and interpretation.  For example, a Dirichlet process model may be used for gene-level intercepts within a regression model for marker probabilities that includes additional prior covariates \citep{lewinger2007}.  Incorporating additional gene-level prior information, such as allowing greater dependence within known functional gene networks \citep{zhang2014}, is also a promising direction of future work.   

Our focus is on association testing.  However, markers that are statistically associated with a given phenotype may not affect the phenotype directly, especially if markers are correlated (e.g., linkage disequilibrium in genetic data).  Our gene-specific model and other prior information  can also be used in the context of model inclusion probabilities, to select markers that have novel predictive power for a given phenotype and are therefore more likely to be causal \citep{wilson2010, zhang2014, duan2013}.  However, computational scaling for high-throughput data is often a challenge in such problems; alternatively, markers identified via association testing may subsequently be included as phenotype predictors in a second stage model \citep{yazdani2015}.

In our context we consider multiple hypotheses, in which the hypotheses are naturally grouped by markers within a gene, but there are similar scenarios in other areas of genomics research.  For example, when screening multiple genes for a phenotypic association (e.g., via microarray or RNA-seq data) the genes can be partitioned into groups based on pathways or other prior information. Frequentist methods that provide appropriate type I error control over genes and gene sets have been developed \citep{Benjamini2008, heller2009, li2014}, and these methods can be generalized to other problems that involve testing hypotheses over multiple sets.  Broadly, our proposed model defines a general prior for multiple hypothesis testing within a Bayesian framework when the hypotheses can be partitioned into sets.

\section*{Acknowledgments}
We thank Dr. Allison Ashley-Koch and Dr. Sandeep Dave for fruitful scientific discussions that motivated this work. This work was supported by the National Institute of Environmental Health Sciences (NIEHS) [R01-ES017436] and National Institutes of Health National Center for Advancing Translational
Sciences (NIH / NCATS) [ULI RR033183 \& KL2 RR0333182].  

\appendix
\section{Posterior Computation}
\label{appendix}

This appendix provides details on posterior computation for the methylation screening application described in Section~ref{app}.

First, we estimate and fix a dictionary of normal kernels truncated between $0$ and $1$, which will be used as mixture components for the density at each CpG site.  These are estimated as described in Section 5 of \citet{lock2015}.  In particular, the number of kernels is determined by out-of-sample cross validation of the log posterior density.  For the present application this yields $K=8$ kernels that appropriately span the data range from $0$ to $1$. 

Let $\Pi_{gm}^{(0)} = (\pi_{gm1}^{(0)},\hdots,\pi_{gmK}^{(0)})$ be the kernel probability weights that define the generative distribution for gene $g$ and site $m$ for group $0$, and let $\Pi_{gm}^{(1)}$ be the kernel probability weights for group $1$.   Under the null model $H_{0gm}$, the mixing weights are the same for both groups:  $ \Pi_{gm}^{(0)}= \Pi_{gm}^{(1)}$.  The kernel weights are assumed to be generated from a Dirichlet($\lambda$) distribution, where $\lambda$ is a hyper-parameter that is inferred during the kernel estimation stage and fixed.  Under $H_{1gm}$, $\Pi_{gm}^{(0)}$ and $\Pi_{gm}^{(1)}$ are considered independent realizations from Dirichlet($\lambda$). 

Under this framework, posterior draws from the gene-level prior model described in Section 3 of the main article are incorporated into Gibbs sampling for the kernel testing parameters.  The full sampling algorithm is described below.    

\begin{enumerate}
\item Draw the kernel that generated each observation $T^{(i)}_{gmn} \in {1,\hdots,K}$ for genes $g=1,\hdots,G$, markers $m= 1,\hdots,M_g$, classes $i=0,1$ and samples $n=1,\hdots,N_i$.   The conditional probability that the given value is realized from component $k$ is 
 \[P(T_{mn}^{(i)}=k\mid X_m^{(i)}, \Pi_m^{(i)}) \propto \pi_{gmk}^{(i)} f(X_{gmn}^{(i)} | \mu_k,\sigma_k,[0,1]),\] 
 where $f(\cdot)$ defines the density of a truncated normal distribution.
\item Designate null markers $H_{0,gm}$ for $g=1,\hdots,G$, $m=1,\hdots,M_g$.  The conditional posterior probability is 
\[P(H_{0,gm} \mid X,Y,p_g)  = \frac{p_g \beta(\lambda)\beta(\vec{n}_m+\lambda)}{p_g \beta(\lambda) \beta(\vec{n}_m+\lambda)+(1-p_g) \beta(\vec{n}_m^{(0)}+\lambda) \beta (\vec{n}_m^{(1)}+\lambda)},\]
where $\vec{n}_{gm}^{(0)} = (n_{gm1}^{(0)},\hdots,n_{gmK}^{(0)})$  is the number of subjects in group $0$ that belong to each kernel $k$ in marker $g,m$, $\vec{n}_{gm}^{(1)}$ is defined similarly for group $1$, and $\vec{n}_{gm}= \vec{n}_{gm}^{(0)}+\vec{n}_{gm}^{(1)}$.  
\item Draw weights $\{\Pi_{gm}^{(0)},\Pi_{gm}^{(1)}\}_{m=1}^M$.   Under $H_{0,gm}$, $\Pi_{gm}^{(0)}=\Pi_{gm}^{(1)} \sim \text{Dirichlet}(\lambda+\vec{n}_{gm})$. Otherwise,  $\Pi_{gm}^{(0)} \sim \text{Dirichlet}(\lambda+\vec{n}_{mg}^{(0)})$ and    $\Pi_{gm}^{(1)} \sim \text{Dirichlet}(\lambda+\vec{n}_{mg}^{(1)})$.
\item Allocate gene-level Dirichlet indices $C_g$ for $g = 1,\hdots, G$:
\[P(C_g=h \mid \theta_\cdot, H_{0,g \cdot}) \propto \pi_h \theta_h^{S_g} (1-\theta_h)^{M_g-S_g}\] 
 where $S_g$ is the number of null markers in gene $g$, $S_g=\sum_{m=1}^{M_g} \mathbbm{1}(H_{0,gm})$.
 \item Update the weights $\pi_h$ for $h=1,\hdots,H$.  First, draw the stick-breaking weights $V_1,\hdots,V_{H-1}$ by 
 \[(V_h \mid C_\cdot) \sim \mbox{Beta} \left(1+\sum_{g=1}^G \mathbbm{1}(C_g=h),\alpha+\sum_{g=1}^G \mathbbm{1}(C_g>h)\right), \]
 with $V_H=1$.  Then set  $\pi_h = V_h \prod_{l<h} (1-V_l)$ for $h=1,\hdots,H$.
 \item Update the atoms $\theta_h$ for $h=1,\hdots,H$:
 \[(\theta_h \mid C_\cdot,H_{0, \cdot \cdot}) \sim \mbox{Beta} \left(a+\tilde{S}_h,b+\tilde{M}_h-\tilde{S}_h\right),\]
 where $\tilde{M}_h$ is the total number of markers in genes allocated to cluster $h$, and $\tilde{S}_h$ is the number of null markers:
\[\tilde{M}_h = \sum_{\{g: C_g=h\}} M_g \, \, \, , \, \, \, \tilde{S}_h = \sum_{\{g: C_g=h\}} S_g.\]
Set $p_g = \theta_{C_g}$ for $g=1,\hdots,G$. 
\end{enumerate}

We use a simple uniform prior for the base distribution of $p_g$ ($a=b=1$).   
 
  \begin{figure}[!ht]
  \centering{\includegraphics[width=0.8\linewidth]{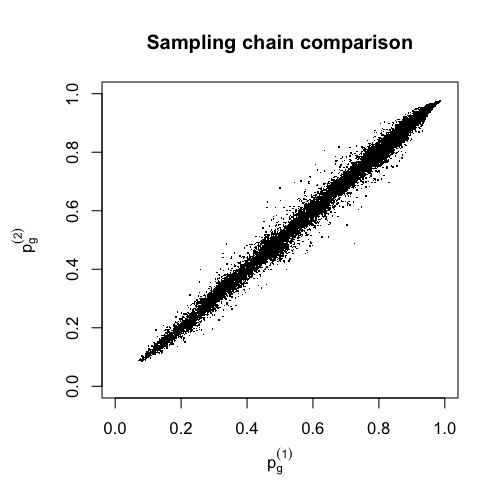}}
\caption{Scatterplot of estimated gene-level prior from two independent sampling chains.}
\label{fig:chaincomp}
\end{figure}

For the high-throughput data considered, computing is not trivial, costing approximately 30 seconds per Gibbs cycle.  However, less than 1\% of computing time is spent on the draws for the gene-level prior parameters (steps 4-6). We find that draws mix well and converge very quickly to a stationary posterior.  We run two parallel chains, with different initializations, for 1000 cycles, with the first 200 treated as burn-in.  Figure~\ref{fig:chaincomp} shows good agreement of estimated gene-level prior probabilities between the two chains.  
 
\newpage 
%\section*{References}
% Either type in your references using
% \begin{thebibliography}{}
% \bibitem{}
% Text
% \end{thebibliography}

  \bibliographystyle{abbrvnat} 
 \bibliography{library2}

\end{document}